\documentstyle[multicol,aps,pre,epsf]{revtex}
\begin{document}
\title{Unusual Dynamical Scaling in the Spatial Distribution of \\ 
Persistent Sites in 1D Potts Models} 
\author{A. J. Bray and S. J. O'Donoghue}
\address{Department of Physics and Astronomy, The University, Manchester,
M13 9PL, United Kingdom}

\date{\today}
\maketitle

\begin{abstract}
The distribution, $n(k,t)$, of the interval sizes, $k$, between clusters 
of persistent sites in the dynamical evolution of the one-dimensional 
$q$-state Potts model is studied using a combination of numerical 
simulations, scaling arguments, and exact analysis. It is shown to have 
the scaling form $n(k,t) = t^{-2z} f(k/t^z)$, with 
$z={\rm max}(1/2,\theta)$, where $\theta(q)$ is the persistence exponent 
which describes the fraction, $P(t) \sim t^{-\theta}$, of sites which 
have not changed their state up to time $t$. When $\theta > 1/2$, the 
scaling length $t^\theta$ for the interval-size distribution is larger 
than the coarsening length scale $t^{1/2}$ that characterizes spatial 
correlations of the Potts variables. 
 
\end{abstract}

\begin{multicols}{2}
\section{Introduction}
The discovery of persistence has recently generated considerable
interest in understanding the statistics of first passage problems in
spatially extended nonequilibrium systems, both theoretically and 
experimentally. The definition of persistence is as follows. Let
$\phi(x,t)$ be a stochastic variable fluctuating in space and time
according to some dynamics. The persistence probability is simply the
probability $P(t)$ that at a fixed point in space, the quantity
${\rm sgn}\,[\phi(x,t)-\langle\phi(x,t)\rangle]$ does not change up to time
$t$. In many systems of physical interest a power law decay, $P(t)\sim 
t^{-\theta}$, is observed, where $\theta$ is the persistence exponent
and is, in general, nontrivial. The nontriviality of $\theta$ emerges
as a consequence of the coupling of the field $\phi(x,t)$ to its
neighbours, since such coupling implies that the stochastic process at
a fixed point in space and time is non-Markovian.

Persistence phenomena have been widely studied in recent years 
\cite{p0,p1,DHP,p2,p3,p4,p5,p6,p7,p8,p9,OCB}. Theoretical and computational 
studies include spin systems in one \cite{p1,DHP} and higher \cite{p2} 
dimensions, diffusion fields \cite{p3}, fluctuating interfaces \cite{p4}, 
phase-ordering dynamics \cite{p5}, and reaction-diffusion systems \cite{p6}. 
Experimental studies include the coarsening dynamics of breath figures 
\cite{p7}, soap froths \cite{p8}, and twisted nematic liquid crystals 
\cite{p9}. Persistence in nonequilibrium critical phenomena has also been 
studied in the context of the global order parameter, $M(t)$, (e.g.\ the 
total magnetization of a ferromagnet), regarded as a stochastic process 
\cite{OCB}. 

In the present work we consider spatially extended systems with a
nonequilibrium field $\phi(x,t)$, which takes discrete values, at each 
lattice site $x$. The field then evolves in time $t$ through interactions 
with its neighbours. The persistence probability at time $t$ is defined as 
the fraction of sites in which the stochastic field $\phi(x,t)$ did not
change its value in the time interval $[0,t]$. The physical interpretation
of $\phi(x,t)$ could be for, for example, the coarsening spin field in
the Ising model after being quenched to low temperature from an
initial high temperature, the sign of a diffusion field starting from a random
initial configuration or the sign of the height, relative to the mean height, 
of a fluctuating interface. As the stochastic field evolves in time, such 
systems develop regions of persistent and nonpersistent sites. Since the 
number of persistent sites decays with time according to 
$P(t) \sim t^{-\theta}$, the persistent clusters shrink in size and number 
with a corresponding growth in the size of the nonpersistent regions.

Recently, Manoj and Ray (MR) \cite{MR} have studied such spatially extended 
systems in one dimension (1D) within the context of the $A+A \rightarrow 0$ 
reaction diffusion model, which is equivalent to the 1D Ising model. They 
found that the length scale which characterises the interval sizes between 
persistent clusters {\em apparently} has a different time-dependence 
from the length scale characterising the walker separations (or spin 
correlations, in the Ising representation), and futhermore seems to depend 
on the initial walker density. If true, this result would be surprising. 
Naively, one would expect the initial walker density to be irrelevant to 
the asymptotic dynamics, and the coarsening length scale, set by the 
mean distance between walkers, to be the relevant length scale for all 
spatial correlations. We shall show that the former expectation is correct, 
but the latter, in general, is not.

In this paper we generalise, and reinterpret, this study in the context 
of the $q$-state Potts model, which has $q$ distinct but equivalent 
ordered phases. The $A+A \rightarrow 0$ model corresponds to the $q=2$ 
Potts model, i.e.\ the Ising model, with the walkers identified as the 
domain walls between up and down Ising spins.  In the $q$-state Potts model 
the random walkers represent domain walls between Potts states. 
At each time step, every walker hops randomly left or right. Persistent 
Potts sites are those at which the Potts state has never changed, i.e.\ 
sites which have never been crossed by a walker. If two walkers occupy 
the same site they either {\em coalesce}, with probability $(q-2)/(q-1)$, 
or {\em annihilate}, with probability $1/(q-1)$, these numbers being the 
probabilities that the states on the farther sides of the walkers are the 
same (annihilation) or different (coalescence). The persistence 
probability (the fraction of sites that have never been jumped over by a 
walker) decays as a power of time, with a $q$-dependent exponent, 
$P(t) \sim t^{-\theta(q)}$. 
 
Among the questions which emerge naturally in such a study are 
(i) What is the dominant length scale in the problem, as far as the 
persistent structures are concerned? 
(ii) What is the nature of the spatial distribution of the 
persistent sites? (iii) What is the average size of a persistent cluster?

An important point to make at the outset concerns the different length 
scales associated with the walkers and with the persistent sites. The 
walker density decays as $t^{-1/2}$ for any $q$, so the mean distance 
between walkers grows as $L_w(t) \sim t^{1/2}$. The density of persistent 
sites decays as $t^{-\theta}$, so the mean distance between these sites 
grows as $L_p(t) \sim t^\theta$. This simple observation immediately 
suggests that the spatial structure of the persistent sites for 
$\theta > 1/2$, where $L_p$ is the larger length scale, will be very 
different from when $\theta < 1/2$, where $L_w$ is the larger length. 
This is precisely what we find: the characteristic length scale, 
$L_{int}$, controlling the distribution of the {\em intervals} 
between clusters of persistent sites, is given by 
$L_{int} = {\rm max}\,(L_w, L_p)$. For the Ising case studied 
earlier \cite{MR}, one has \cite{DHP} $\theta = 3/8$, which is 
`close' to $1/2$. We believe the proximity of these two exponents is 
responsible for the apparently non-universal behavior in the Ising 
system reported in MR. 

Derrida et al.\ \cite{DHP} have obtained an exact 
expression for $\theta$ for 1D Potts models with arbitrary $q$:
\begin{equation}
\theta(q) = -\frac{1}{8} + \frac{2}{\pi^2}\,
\left[\cos^{-1}\left(\frac{2-q}{\sqrt{2}\,q}\right)\right]^2\ .
\label{theta(q)} 
\end{equation}
The value of $q$ corresponding to $\theta = 1/2$ is 
$q_c = 2/[1+\sqrt{2}\cos(\sqrt{5}\pi/4)] = 2.70528\ldots$, 
so $\theta(q) > 1/2$ for all integer $q \ge 3$. Note that the 
probabilistic algorithm for implementing the Potts model through 
the annihilation or coalescence of random walkers (domain walls) allows 
$q$ to be a real number, $q \ge 2$, while an equivalent Ising spin 
representation (see section IV) of the Potts persistence problem  
even allows $1<q<2$! By this means means we can explore a range of $\theta$  
above and below 1/2. 

Our main result is that the scale length controlling the distribution of 
interval sizes between persistent clusters is given by $t^{1/2}$ when  
$\theta < 1/2$, but by $t^\theta$ when $\theta > 1/2$. We find no evidence  
for any dependence of the asymptotic scaling distribution on the initial 
walker density (other than through nonuniversal amplitudes). 

This paper is organised in the following manner. In section II we outline 
a scaling phenomenology within which the above questions can be addressed. 
In section III we give exact results for the $q = \infty$ Potts model. 
Finally, in section IV, we present extensive numerical simulations for the
$q$-state Potts model which confirm our predictions based on general 
scaling arguments and the exact large-$q$ results. Section V concludes 
with a discussion and summary of the results.

\section{Scaling Phenomenology}
The basic scaling phenomenology was introduced by Manoj and Ray, although 
we will adopt a slightly different notation. We are interested in the 
distribution of the sizes of the {\em intervals} between persistent sites. 
We define an interval size, $k$, as the number of {\em non-persistent} 
sites between two consecutive persistent sites. If the interval size is 
zero, the sites belong to the same {\em cluster} of persistent sites, and 
we henceforth consider only intervals of non-zero size. We define $n(k,t)$ 
to be the number of intervals (per site) of size $k$ at time $t$. A 
natural dynamical scaling assumption is 
\begin{equation}
n(k,t) = t^{-2z}\,f(k/t^z)\ ,
\label{scaling}
\end{equation} 
where $L(t)=t^z$ is the `characteristic' length scale at time 
$t$. (A dynamical exponent $z$ is conventionally defined via 
$L(t) \sim t^{1/z}$, rather than $t^z$. Here we are following 
the notation of MR). 

The rationale for the scaling form (\ref{scaling}) is as follows. 
The number of non-persistent sites (per site) is given by 
\begin{equation}
Q(t) = \sum_{k=1}^{\infty} k\,n(k,t)\ .
\label{np} 
\end{equation}
But since the number of persistent sites (per site), $P(t)$, decays to 
zero as $P(t) \sim t^{-\theta}$, and $P(t)+Q(t)=1$, it follows that 
$Q(t) \to 1$ for $t \to \infty$. Converting the sum (\ref{np}) to an 
integral, with lower limit zero (valid as $t\to \infty$ provided the 
integral converges), we see that the prefactor $t^{-2z}$ in (\ref{scaling}) 
is precisely what is needed to satisfy the required condition 
$Q(t) \to {\rm const}$ for $t \to \infty$. 

Consider next the number, $N_c(t)$, of persistent clusters per site. 
Since the number of clusters is equal to the number of intervals, $N_c$ 
is given by
\begin{equation}
N_c(t) = \sum_{k=1}^\infty n(k,t). 
\label{cl}
\end{equation}
Converting the sum to an integral (with lower limit zero), using 
(\ref{scaling}) and assuming the integral converges, gives 
$N_c \sim t^{-z}$. Thus the mean distance between persistent clusters, 
i.e.\ the mean interval size, increases as $t^z$. 

Is this reasonable? Let us recall that there are two length scales in 
the system, $L_w \sim t^{1/2}$, the mean distance between walkers, and 
$L_p \sim t^\theta$, the mean distance between persistent sites. 
To make further progress we make the following two assumptions, both of which 
are confirmed by our numerical studies, by exact results for $q=\infty$, and 
by heuristic arguments to be expounded below: 
(i) The mean cluster size tends to a constant for $t \to \infty$, and 
(ii) The length scale $L_{int} = t^z$ that controls the interval size 
distribution is the {\em larger} of $L_w$ and $L_p$, i.e.\ 
$z = {\rm max}\,(1/2, \theta)$. 

From assumption (i) we deduce that $N_c \sim t^{-\theta}$, i.e.\ $z=\theta$, 
which is consistent with assumption (ii) provided $\theta > 1/2$. What if 
$\theta < 1/2$? Then we still require $N_c \sim t^{-\theta}$, but now 
$z=1/2$, so the result $N_c \sim t^{-z}$, derived from (\ref{cl}) breaks 
down. Going back to (\ref{cl}), we infer that this breakdown requires 
that the conversion of the sum to an integral be invalid -- the sum must 
have its dominant contribution from $k$ of order unity, rather than $k$ of 
order $t^z$. This in turn requires that the scaling function $f(x)$ in 
(\ref{scaling}) have a singular small-$x$ limit of the form \cite{MR} 
\begin{equation}
f(x) \sim x^{-\tau},\ \ \ x \to 0\ ,
\end{equation}
with $1 < \tau < 2$. Using this form in (\ref{scaling}), 
with $z=1/2$, and inserting the result into (\ref{cl}), gives 
$N_c \sim t^{-(1-\tau/2)}$. Comparing this with $N_c \sim t^{-\theta}$ 
fixes the value of $\tau$:
\begin{equation}
\tau = 2(1-\theta),\ \ \ \ \theta < 1/2\ .
\end{equation}

\begin{figure}
\narrowtext
\centerline{\epsfxsize\columnwidth\epsfbox{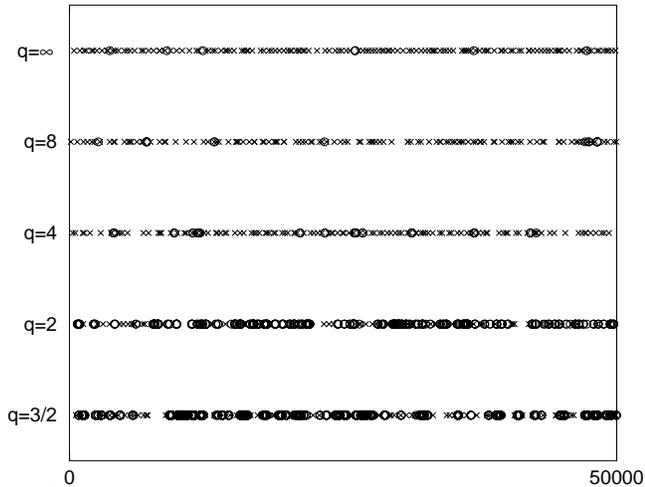}}
\caption{Snapshots showing the relative densities of persistent sites 
(0) and random walkers (x) for various $q$, at time $t=10^4$. 
The Ising representation was used to generate the $q=3/2$ data, with 
the Potts representation used for the other $q$-values. 
} 
\label{Fig0}
\end{figure}

We first present a heuristic argument for the result 
$z={\rm max}\,(1/2, \theta)$. Consider first the case $\theta > 1/2$. 
In this regime, the distance between clusters is much greater that the 
distance between walkers, i.e.\ there are, on average, many 
($\sim t^{\theta-1/2}$) random walkers between each consecutive pair 
of persistent clusters, as illustrated in Figure \ref{Fig0}. 
However, walkers separated by distances large compared to $t^{1/2}$ 
are essentially uncorrelated, because any correlations are mediated 
by the random walkers, and correlations between 
walkers decay on the length scale $t^{1/2}$. Therefore one expects the 
intervals between clusters, measured on the scale $t^\theta$,  to be 
independent random variables, and the interval size distribution to scale 
with this length. This argument also suggests that the locations of the 
clusters as a function of position on the lattice are described by a 
Poisson process, i.e.\ the interval size distribution is an exponential,  
$n(k,t) = \langle k \rangle \exp(-k/\langle k \rangle)$, with 
$\langle k \rangle \sim t^\theta$, for $\theta > 1/2$. These expectations 
are borne out by our numerical studies. 

For $\theta < 1/2$, the opposite is true: there are many clusters between 
each pair of walkers (see the $q=3/2$ snapshot in Figure \ref{Fig0}). 
The largest length scale is set by the walker spacing, $L_w \sim t^{1/2}$, 
and the interval size distribution will scale with this length. On 
smaller scales, there will be residual structure in the cluster 
distribution left from an earlier epoch when the number of walkers was 
larger. Indeed, on these smaller scales the persistent sites form a fractal 
set \cite{fractal,MR2,Jain}. Consider two sites separated by a distance $r$. 
The probability, $P_2(r,t)$, that {\em both} are persistent has the scaling 
form, 
\begin{equation}
P_2(r,t) = t^{-2\theta}\,F(r/t^{1/2})\ ,
\label{both}
\end{equation}
where $F(x) \to {\rm const}$ for $x \to \infty$, since for 
$r \gg t^{1/2}$ the sites are uncorrelated. The residual structure 
remaining on scales $r \ll t^{1/2}$, however, means that if the first 
site is persistent, which occurs with probability of 
order $t^{-\theta}$, the probability that the second one is also persistent 
depends only on $r$. This implies $F(x) \sim x^{-2\theta}$ for $x \to 0$. 
The number of persistent sites within a distance $R$ of a given persistent 
site can therefore be estimated as $\int_0^R dr\, r^{-2\theta} \sim R^{d_f}$, 
where 
\begin{equation}
d_f = 1-2\theta 
\label{df}
\end{equation}
is the fractal dimension of the persistent set. 

Clearly this result only makes sense for $\theta < 1/2$, since $d_f$ 
cannot be negative. This suggests $d_f = 0$ for $\theta \ge 1/2$, 
corresponding to point-like objects, i.e.\ isolated finite clusters. 
For $q=\infty$, for example, where $\theta = 1$, we find that there 
is typically one cluster of persistent sites, with a fixed mean size, 
per scale length $L_p \sim t$. 
  
We now present a heuristic argument in support of our claim that the mean 
cluster size, $\langle l \rangle$,  approaches a constant as $t \to \infty$ 
for all values of $\theta$. The initial steps of the argument follow the 
approach of MR. The total number of clusters per site can be written as  
\begin{equation}
N_c(t) = P(t) - P_w(t),
\end{equation}
where $P_w(t)$ is the fraction of walker sites which have never been visited 
by a walker, i.e.\ the persistence probability for walker sites. Such sites 
form `spacers' between each pair of persistent Potts states in a cluster, 
there being exactly one more persistent Potts site than persistent walker 
site in each cluster. The mean cluster size is 
\begin{equation}
\langle l \rangle = \frac{P(t)}{N_c(t)} 
= \left[1 - \frac{P_w(t)}{P(t)}\right]^{-1}\ .
\label{lav}
\end{equation}
We expect, on universality grounds, that the exponents describing 
the decay of $P(t)$ and $P_w(t)$ should be the same (i.e.\ $\theta$), 
but the (nonuniversal) amplitudes will be different, 
i.e.\ $P(t) \to At^{-\theta}$, while $P_w(t) \to A_wt^{-\theta}$, 
with $A_w < A$. Inserting these forms in (\ref{lav}) gives the 
limiting value of the mean cluster size as 
$\langle l \rangle_\infty = (1 - A_w/A)^{-1}$. 
This number is nonuniversal and is determined by the 
initial distribution of cluster sizes.  

\section{Exact Results for $A+A \to A$}
Many of the general features of the dynamics for $\theta>1/2$ are 
exemplified by the $q \to \infty$ limit, for which the walker dynamics 
reduces to $A + A \to A$, i.e.\ the walkers always aggregate on contact. 
At $t=0$ all sites are persistent. The random walkers initially present 
divide the sites into clusters of persistent sites. Clearly no cluster 
can increase in size. Consider a cluster of initial size $l_0$. 
We first calculate the probability density, $p_{l_0}(l,t)$, that the cluster 
survives to time $t$ and has size $l$. The key point is that, 
for $A + A \to A$ dynamics, we need only consider the two walkers on 
either side of the initial cluster, since subsequent coalescence 
processes do not affect the random walk dynamics of these two walkers. 
The cluster and the two walkers can therefore be treated in isolation. 

For simplicity we begin by treating the continuum limit in which the 
walkers are regarded as continuous-time random walkers on a continuous 
space. This should be correct in the limit $l_0 \gg 1$, an assertion we 
can subsequently test. Let the ends of the initial cluster (and the two 
walkers) be located at $x=0$ and $x=l_0$. Each walker obeys an equation of 
the form $dx/dt = \xi(t)$, where $\xi(t)$ is a Gaussian white noise with 
mean zero and correlator $\langle \xi(t)\xi(t') \rangle = 2\delta(t-t')$. 
We first write down the probability distribution $P_r(x_r,t)$ of the 
rightmost excursion, $x_r$, up to time $t$, of the left walker. An 
elementary calculation gives
\begin{equation}
P_r(x_r,t) = \frac{1}{\sqrt{\pi t}}\,\exp\left(-\frac{x_r^2}{4t}\right)\ .
\label{P_r}
\end{equation}
For the cluster to survive, we require $x_r < l_0$, so in the limit 
$t \gg l_0^2$ the exponential factor in (\ref{P_r}) can be dropped. 
The probability distribution, $P_l(x_l,t)$, of the leftmost excursion 
of the right walker (from its initial position) is given by a 
similar expression. Clearly $l = l_0 - x_l - x_r$ is the residual size of 
the cluster at time $t$, where $l \le 0$ means the cluster has disappeared.
The probability density $p_{l_0}(l,t)$ that the cluster has survived and has 
size $l$ is therefore given, for $t \gg l_0^2$, by 
\begin{eqnarray}
p_{l_0}(l,t) & = & \int_0^{l_0-l} \frac{dx_l}{\sqrt{\pi t}} 
\int_0^{l_0-l} \frac{dx_r}{\sqrt{\pi t}}\,\delta(l_0 - l - x_l - x_r) 
\nonumber \\   
& = &\frac{(l_0-l)}{\pi t}\ .
\label{pl0} 
\end{eqnarray}
The probability, $p_{surv}(l_0,t)$, that the cluster survives till time $t$ 
is given by
\begin{equation}
p_{surv}(l_0,t) = \int_0^{l_0} dl \frac{(l_0-l)}{\pi t} 
                = \frac{l_0^2}{2\pi t}\ .
\end{equation}

Immediately we see that the mean interval between surviving clusters 
grows as $t$ ($=t^\theta$, since $\theta(q=\infty)=1$), which is much 
larger than the mean interval between walkers, which grows as $t^{1/2}$.

The mean length of surviving clusters, of given initial size $l_0$, 
in the $t \to \infty$ limit is
\begin{equation}
\langle l \rangle_\infty = \frac{2\pi t}{l_0^2}
\int_0^{l_0}dl\,l\,\frac{(l_0-l)}{\pi t} = \frac{1}{3}\,l_0\ .
\end{equation}
If the initial clusters have a distribution of sizes, a straightforward 
calculation gives $\langle l \rangle_\infty = 
\langle l_0^3 \rangle_0/3\langle l_0^2 \rangle_0$, where 
$\langle \ldots \rangle_0$ indicates an average over this distribution, 
while the fraction of clusters which survive is 
$\langle l_0^2 \rangle_0/2\pi t$.

The above calculations demonstrate that the mean cluster size approaches 
a constant at late times, and that the scale length for the sizes of the 
intervals between clusters grows as $t$, and not as the naive scaling 
length $t^{1/2}$ associated with the underlying domain wall coarsening. 
In fact we can calculate the interval size distribution, $n(k,t)$, 
explicitly for this model. First recall that the surviving clusters are 
separated, at late times, by many ($\sim t^{1/2}$) walkers. The fate of 
a given cluster depends only on the nearest walker on either side. The 
motion of these walkers is uncorrelated with that of the walkers 
bordering  other clusters, because the correlation length for walkers 
grows only as $t^{1/2}$. Therefore we can assume, in the scaling limit 
$k \to \infty$, $t \to \infty$, with $k/t$ fixed, that clusters survive 
{\em independently} with probability $l_0^2/2\pi t$ (where we have 
specialized to initial clusters of fixed size $l_0$). The probability 
distribution of the interval size $k$ between neighboring clusters is 
therefore exponential, with mean $\langle k \rangle = 2\pi t/l_0^2$, 
while $n(k,t) = \langle k \rangle^{-2}\exp(-k/\langle k \rangle)$, which 
has the scaling form (\ref{scaling}) with $z=1$. This exponential form 
is in excellent agreement with simulation data presented in section IV. 

Since the initial clusters used in the $q=\infty$ simulations are quite 
small (2 or 4), the continuum limit is not expected to be quantitatively 
correct. To conclude this section, therefore, we quote the asymptotic 
mean cluster size for clusters of arbitrary initial size $l_0$.  
The details are given in the Appendix: the result is 
$\langle l \rangle_\infty  = (l_0+2)/3$, which generalizes the continuum 
result $\langle l \rangle_\infty = l_0/3$. 

\section{Numerical Simulations}
Let us recall the zero-temperature coarsening dynamics of the 1D
$q$-state Potts model, starting from a random initial configuration. 
To maximize the speed of the program, we adopt two-sublattice parallel 
updating. The zero-temperature dynamics proceeds as follows: in each 
time step every spin on one of the sublattices changes its colour to 
that of one of its two neighbours with equal probability, the two 
sublattices being updated alternately. The dynamics of such systems 
can equivalently be formulated in terms of the motions of the domain 
walls as a reaction-diffusion model. The domain walls can be
considered as random walkers performing independent random walks. 
Whenever two walkers meet, they annihilate ($A+A \rightarrow 0$) with 
probability $1/(q-1)$ or aggregate ($A+A \rightarrow A$), to
become a single walker, with probability $(q-2)/(q-1)$. The persistence 
is the probability that a fixed point in space has not been traversed
by any random walker. For $q=2$, the particles only annihilate and
hence this is equivalent to the Glauber model, whereas in the 
$q=\infty$ limit the walkers only aggregate. 

This algorithm is, however, restricted to modelling $q$-state Potts 
models with $q \ge 2$, since $q<2$ generates negative probabilities, 
while the general result (\ref{theta(q)}) allows any real $q \ge 1$.  
In order to simulate values of $q$ in the interval $(1,2)$ we can map 
the Potts model (as far as persistence properties are concerned) on to 
an Ising representation with a fraction $1/q$ of the sites initially 
pointing up, and initially persistent, while the down spins are 
nonpersistent. Studying the persistence of the minority (majority) spins 
then corresponds to the case $q>2$ ($q<2$) \cite{ben-naim}. The domain 
walls between the up and down spins behave as random walkers, annihilating 
each other on contact with probability one. As far as the persistence 
is concerned, the dynamics of the Potts and Ising models are completely 
equivalent \cite{ben-naim}. For $q>2$, results obtained from both types of 
simulation are entirely consistent.

The numerical simulations are performed on a 1D lattice of size 
$2 \times 10^6$ with periodic boundary conditions. The Ising 
spins, or Potts variables, occupy the odd sites of the lattice,  
while the random walkers are restricted to even-numbered sites. 
Therefore the effective size of the lattice, in terms of Ising/Potts 
spins, is $10^6$. Each walker jumps to an  adjacent even-numbered 
site with equal probability, turning any persistent site it hops over 
into a nonpersistent site. The positions of all walkers 
are updated simultaneously at each Monte-Carlo step. The initial 
configurations are chosen to eliminate the possibility of any `crossover' 
of random walkers when they jump. This can be done by placing them only 
on sites whose positions on the lattice are of the form $4k$, where $k$ is 
an integer. The walkers then occupy subsets of the sites $4k$ and $4k+2$ 
alternately. 

For the Potts simulations ($q>2$), the lattice is initially completely 
persistent. The random walkers are periodically laid down, in clusters 
of uniform initial size $l_0$. The walkers then execute random walks, 
aggregating or annihilating according to the prescribed probabilities, 
$(q-2)/(q-1)$ and $1/(q-1)$ respectively. The simulations are performed 
for initial cluster sizes of 2 and 4. The values of $q$ chosen are 
$q=\infty,8$ and 4, for which the corresponding values of $\theta$ 
are $\theta(\infty)=1$, $\theta(8) \simeq 0.7942$, and 
$\theta(4) \simeq 0.6315$. 

For $q <2$ the Ising representation is employed, in which persistent 
sites (up spins) are periodically laid down on the lattice, with a 
fraction $1/q$ of all sites persistent, and the domain walls execute 
random walks annihilating according to $A+A \rightarrow 0$. The 
value $q=3/2$ was chosen, for which $\theta \simeq 0.2350$. 
The simulations were performed for initial cluster sizes 4 and 8. 
The Ising simulations were also run for $q = 4$ and $8$. The results 
for scaling functions are consistent with the Potts data, but the 
statistics are poorer due to the smaller number of clusters generated. 
For this reason, the data presented in Figs.\ 3-9 were generated  
from the Potts runs. All the simulations (Potts and Ising) were run for 
$10\,000$ Monte-Carlo steps, and the results averaged over 30 
independent runs.

\begin{figure}
\narrowtext
\centerline{\epsfxsize\columnwidth\epsfbox{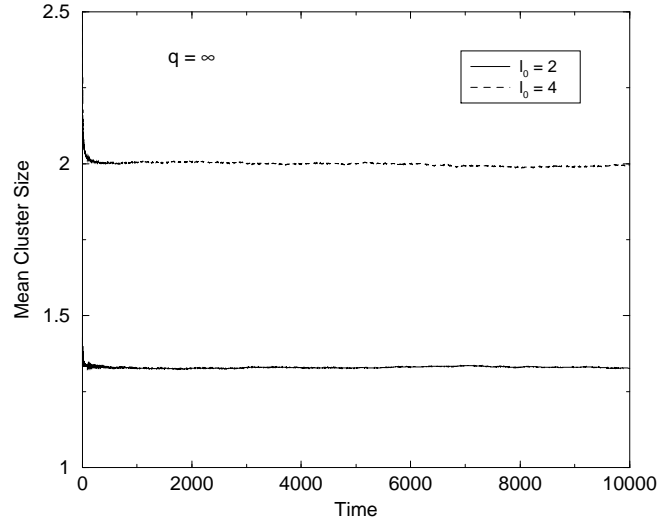}}
\caption{Mean cluster size as a function of time for $q=\infty$ and 
initial cluster sizes $l_0=2$ and 4. 
} 
\label{Fig1}
\end{figure}

The simulations investigate the limiting values of the average 
sizes of the persistent clusters, and the distribution, $n(k,t)$, of 
intervals between persistent clusters with emphasis on its dynamic 
scaling form.  While the first of these lends itself readily to 
direct measurement, the dynamic scaling form for $n(k,t)$ manifests 
itself most clearly in the data through a study of the (complement 
of the) cumulative distribution. We define 
\begin{equation}
I(k,t) = \sum_{k'\ge k} n(k',t)\ .
\end{equation}
Inserting the scaling form (\ref{scaling}) and converting the sum to an 
integral in the scaling limit $k \to \infty$, $t \to \infty$ with 
$k/t^z$ fixed but arbitrary, gives 
\begin{eqnarray}
I(k,t) & = & t^{-2z}\int_k^\infty dk'\,n(k',t) \nonumber \\
       & = & t^{-z}\int_{k/t^z}^\infty dx\,f(x) = t^{-z}\,g(k/t^z)\ .
\label{newscaling}
\end{eqnarray}

\begin{figure}
\narrowtext
\centerline{\epsfxsize\columnwidth\epsfbox{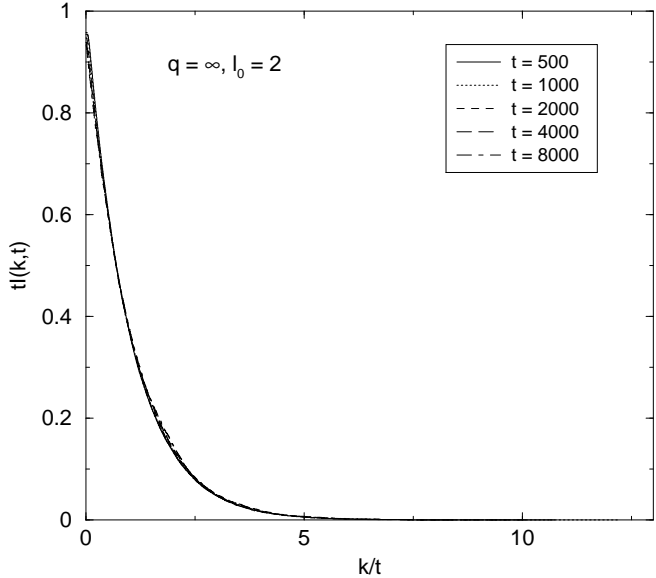}}
\caption{Scaling plot for $q=\infty$, for initial cluster size $l_0=2$. 
} 
\label{Fig2}
\end{figure}

We now discuss the numerical results for the $q=\infty$ model. 
In Figure \ref{Fig1} we plot the mean cluster size against time for  
initial cluster sizes of 2 and 4. The numerical results are summarized 
in Table 1. The numerics clearly support the exact results given in
section III [and derived in the Appendix as equation(\ref{l_av})],  
namely $\langle l \rangle_\infty = (l_0+2)/3$ where $l_0$ is the 
initial cluster size. In order to investigate the dynamic scaling form 
(\ref{newscaling}) for $q=\infty$, we plot, in Figure \ref{Fig2},  
$tI(k,t)$ against $k/t$ for $t=$ 500, 1000, 2000, 4000, 8000, for initial
cluster size 2, where we have used $z=\theta(\infty)=1$ for 
$q=\infty$. Excellent data collapse occurs in agreement with 
the dynamic scaling form (\ref{newscaling}).  

\begin{center}
\begin{tabular}{|c|c|c|}
\hline
  $q$     &       $l_0$   & $\langle l\rangle_\infty$ \\
\hline
$\infty$  &  2(P) & $1.330 \pm 0.003$ \\
          &  4(P) & $1.998 \pm 0.006$ \\  
\hline
  8       &  2(P) & $1.367 \pm 0.002$ \\
          &  2(I) & $1.333 \pm 0.005$ \\
          &  4(I) & $2.018 \pm 0.013$ \\
\hline
  4       &  2(P) & $1.3945 \pm 0.0007$ \\
          &  4(P) & $2.152 \pm 0.001$ \\
          &  4(I) & $2.089 \pm 0.004$ \\
\hline
 1.5      &  4(I) & $2.5880 \pm 0.0002$ \\
          &  8(I) & $4.572 \pm 0.002$ \\
\hline
\end{tabular}
\end{center}

\noindent\underline{Table 1}. Average cluster size at late times, 
$\langle l\rangle_\infty$, for various values of $q$ and various 
initial cluster sizes $l_0$, where P or I indicate whether Potts or 
Ising representations were employed. In all cases the data were 
averaged over times between 1000 and 9000 Monte Carlo steps. 

\bigskip

\begin{figure}
\narrowtext
\centerline{\epsfxsize\columnwidth\epsfbox{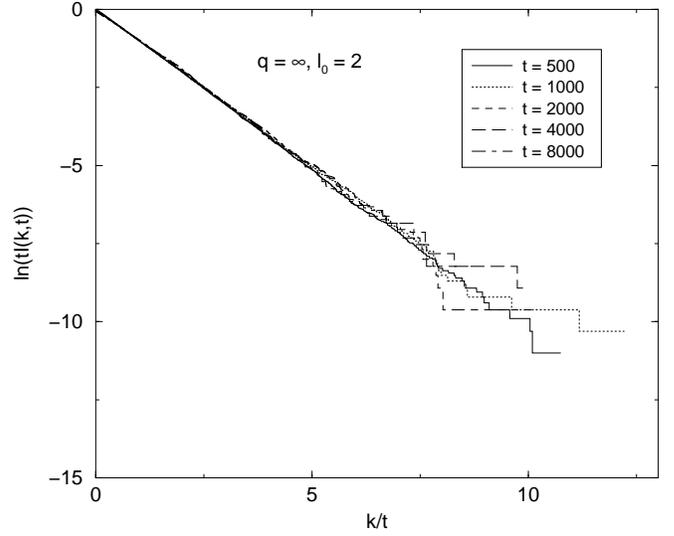}}
\caption{Same as Figure \ref{Fig2}, but presented as a log-linear 
plot to reveal the exponential form of the scaling function. 
} 
\label{Fig3}
\end{figure}

In section III we argued that the scaling function $f(x)$ in 
(\ref{scaling}) should be a simple exponential, which implies 
that $g(x)$ in (\ref{newscaling}) is also a simple exponential. 
To test this prediction, we plot $\ln(tI(k,t))$ against $k/t$ in 
Figure \ref{Fig3}. The data lie on the expected straight line except 
at the latest times, where the deviations from the line are scatter 
due to statistical noise.  Very similar results are obtained for 
initial cluster size 4, confirming that the scaling function $f(x)$ 
is independent of the initial walker density.

\begin{figure}
\narrowtext
\centerline{\epsfxsize\columnwidth\epsfbox{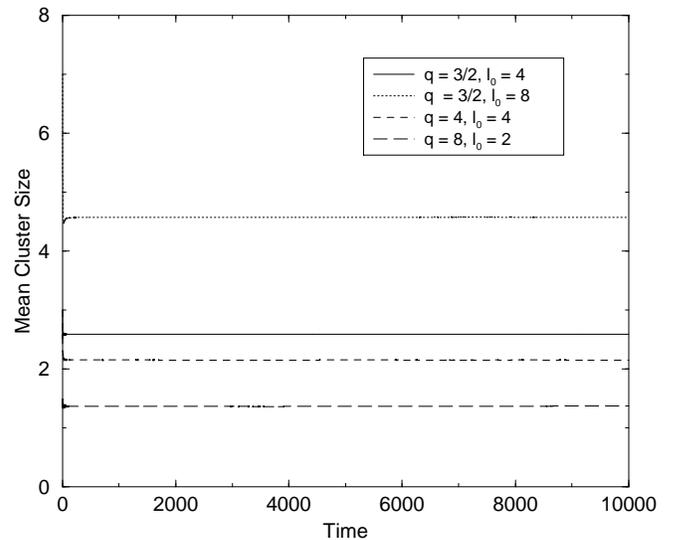}}
\caption{Mean cluster size as a function of time for various 
values of $q$ and various initial cluster sizes. The Potts  
representation was used for $q>2$. 
} 
\label{Fig4}
\end{figure}

\begin{figure}
\narrowtext
\centerline{\epsfxsize\columnwidth\epsfbox{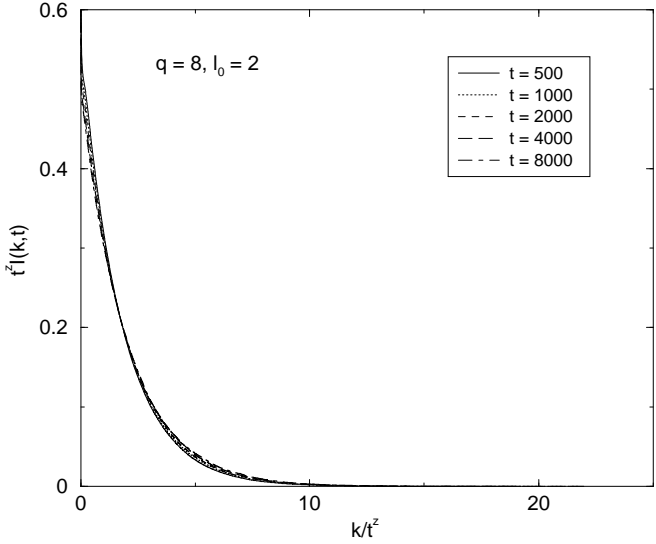}}
\caption{Scaling plot for $q=8$ and initial cluster size $l_0=2$, 
with $z=\theta(8) \simeq 0.7942$. 
} 
\label{Fig5}
\end{figure}

\begin{figure}
\narrowtext
\centerline{\epsfxsize\columnwidth\epsfbox{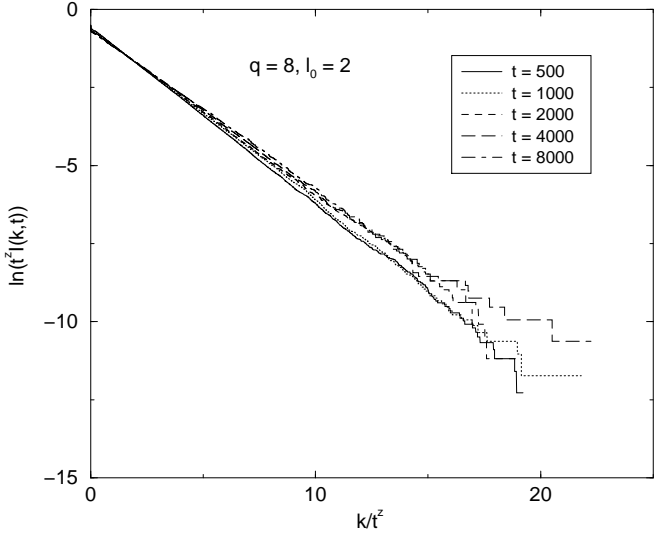}}
\caption{Same as Figure \ref{Fig5}, but presented as a 
log-linear plot to reveal the underlying exponential form. 
} 
\label{Fig6}
\end{figure}

In Figure \ref{Fig4} the mean cluster size is plotted against time 
for various finite values of $q$ and initial cluster size $l_0$. The 
numerical results indicate that the average cluster size tends to a 
constant (see Table I) in all cases, as expected on the basis of the 
heuristic argument in section II. For $q=8$, 
$\theta(q) \simeq 0.79 > 1/2$ and the dominant length scale is still 
determined by the mean distance between persistent clusters. 
We therefore anticipate a scaling form 
of the kind given by (\ref{newscaling}) with $z = \theta(8) $ . 
The scaling plot presented in Figure \ref{Fig5} shows good data 
collapse with this value of $z$, while Figure \ref{Fig6} indicates 
that $g(x)$, and therefore $f(x)$, is again a simple exponential. 
The data collapse is not as good as for $q=\infty$, which we attribute 
to a poorer separation of the length scales $L_p$ and $L_w$ at the 
time scales available. This problem becomes more acute for $q=4$ 
(see the discussion below). 

\bigskip

\bigskip

\begin{figure}
\narrowtext
\centerline{\epsfxsize\columnwidth\epsfbox{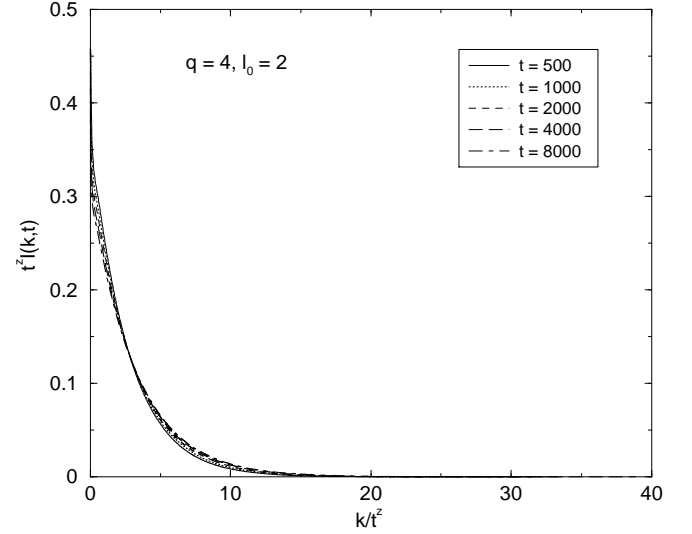}}
\caption{Same as Figure \ref{Fig5}, but for $q=4$, 
with $z=\theta(4)\simeq 0.6315$.
} 
\label{Fig7}
\end{figure}

For $q=4$, $\theta(q) \simeq 0.63 > 1/2$, so we plot (Figure \ref{Fig7}) 
$t^zI(k,t)$ against $k/t^z$ with $z=\theta(4)$. This time the 
scaling  collapse is definitely not as good as for $q=\infty$ and $q=8$, 
reflecting an apparent deviation from the dynamic scaling form 
(\ref{newscaling}). We believe that this deviation reflects the 
proximity of the length scales $L_p \sim t^{0.63}$ and 
$L_w \sim t^{1/2}$, which are not sufficiently well separated on 
the timescales achievable in the simulation, and would disappear 
for asymptotically large times. For example, $L_p/L_w \sim t^{0.1315} 
\simeq 3.26$ at $t=8\,000$, while the corresponding ratio is 
about 14 for $q=8$ and 90 for $q=\infty$. A similar effect may 
explain the apparent deviations from universality (e.g.\ an apparent 
dependence of $z$ on the initial walker density) in the $q=2$ 
simulations of MR. In this case $\theta = 3/8$, and $L_p/L_w 
\sim t^{-1/8} \sim 0.31$ at $t=12\,000$, the largest time reached 
by the MR simulations. So in this case also there is not a strong 
separation of length scales at the times achieved in the simulations 
(a point recently emphasized by Manoj and Ray \cite{MR2}).

\begin{figure}
\narrowtext
\centerline{\epsfxsize\columnwidth\epsfbox{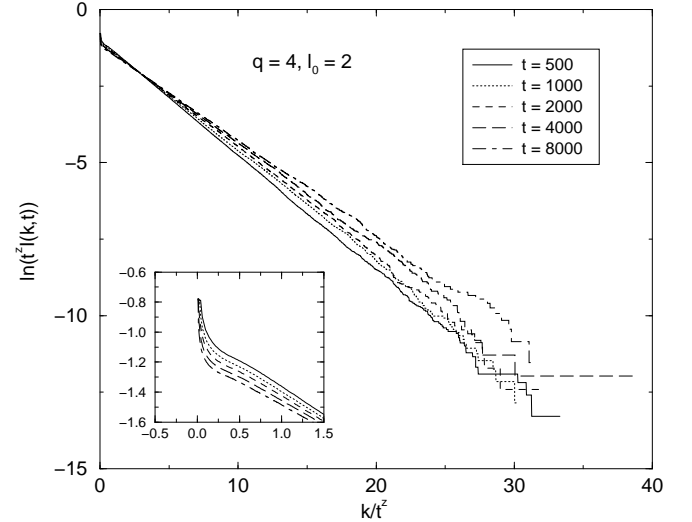}}
\caption{Same as Figure \ref{Fig7}, but presented as a log-linear 
plot to test the predicted exponential form of $g(x)$. The inset 
is an expanded version of the extreme left of the plot.   
} 
\label{Fig8}
\end{figure}

Figure \ref{Fig8} gives the log-linear plot for $q=4$. Again, the data are 
consistent with an exponential scaling function, except at small
scaling variable where there is a perceptible deviation from a straight 
line. In this region, however, the data also do not scale at all well 
either (see inset in Figure \ref{Fig8}), which may be another 
manifestation of the poor separation of length scales: the scaling 
limit requires $k \to \infty$, $t \to \infty$ with $k/t^z$ fixed. 
In particular the condition $k \gg L_w \sim t^{1/2}$,  
which should be satisfied for good scaling, is violated at the small 
values of $k/t^z$ where the upturn in the data occurs. We will return to 
this point in the Discussion. 

\begin{figure}
\narrowtext
\centerline{\epsfxsize\columnwidth\epsfbox{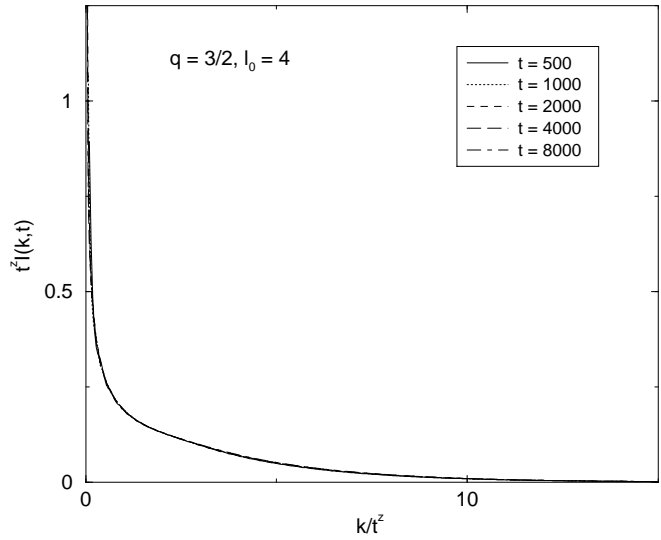}}
\caption{Same as Figure \ref{Fig5}, but for $q=3/2$, with $z=1/2$. 
} 
\label{Fig9}
\end{figure}

To investigate the spatial distribution of the sites in the regime
$\theta < 1/2$ we study $q=3/2$, for which $\theta(q) \simeq 0.235$. 
In Figure \ref{Fig4} we show that the mean cluster size is a constant 
for initial cluster sizes of 4 and 8. In the regime $\theta < 1/2$ 
the density of the persistent sites decays more rapidly than that of 
the walkers. The dominant length scale is therefore no longer given   
by the persistence length, $L_p \sim t^\theta$, but is given instead 
by the mean separation, $L_w \sim t^{1/2}$, of the the walkers. 
Hence we expect asymptotic dynamic scaling of the form (\ref{scaling}) 
and (\ref{newscaling}) with $z=1/2$. In Figure \ref{Fig9} we plot 
$t^{1/2}I(k,t)$ against $k/t^{1/2}$ for an initial cluster size $l_0=4$. 
The numerical results give excellent data collapse, supporting the 
proposed dynamic scaling form. Apart from a change of scale, very 
similar results are obtained for $l_0=8$, supporting the universality 
of the scaling function, i.e.\ the independence of $g(x)$ from the 
initial walker density. We choose $q=3/2$ rather than $q=2$ in order that 
the length scales $L_p$ and $L_w$ be reasonably well separated at late 
times: $L_p/L_w \sim t^{-0.265} \simeq 0.09$ at $t=8\,000$. 

\begin{figure}
\narrowtext
\centerline{\epsfxsize\columnwidth\epsfbox{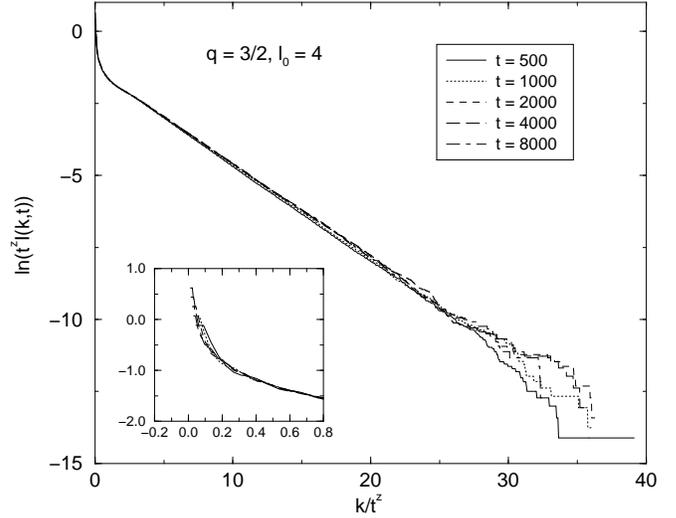}}
\caption{Same as Figure \ref{Fig6}, but for $q=3/2$, and with $z=1/2$. 
The inset shows an expanded version of the extreme left of the plot, 
and suggests a singularity for small $x=k/t^{1/2}$.    
} 
\label{Fig10}
\end{figure}

\begin{figure}
\narrowtext
\centerline{\epsfxsize\columnwidth\epsfbox{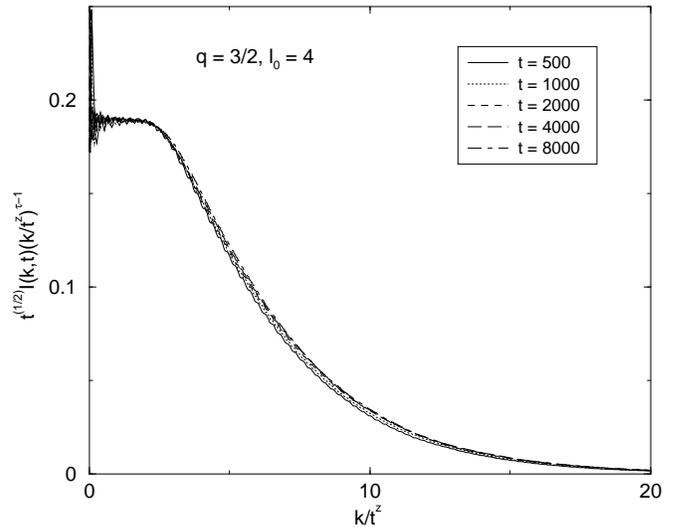}}
\caption{Same data as Figure \ref{Fig9}, but replotted with ordinate  
$t^{1/2}[k/t^{1/2}]^{1-2\theta}I(k,t)$ to show the small-$x$ 
behavior ($x=k/t^{1/2}$) more clearly.     
} 
\label{Fig11}
\end{figure}

Figure \ref{Fig9} reveals a pronounced upturn at small $k$, 
which can be seen more clearly in the log-linear plot of Figure 
\ref{Fig10} and its inset, suggesting a divergence for $k \to 0$. 
This is, in fact, expected from the analysis of section II, which 
predicts, for $\theta < 1/2$, $f(x) \sim x^{-\tau}$ for $x \to 0$, 
with $\tau = 2(1-\theta)$, i.e.\ $\tau$ is in the range $1 < \tau < 2$. 
The scaling function $g(x)$ is given by $g(x) = \int_x^\infty dy\,f(y)
\sim x^{-(\tau-1)}$ for $x \to 0$, i.e.\ $g(x) \sim x^{-(1-2\theta)} 
= x^{-d_f}$, where $d_f = 1-2\theta$ is the fractal dimension of the 
persistent set on scales smaller than $L_w$ [equation (\ref{df})].
According to this prediction, the product $x^{1-2\theta}g(x)$ should 
approach a constant at small $x$. This product is shown in Figure 
\ref{Fig11}. The small-$k$ divergence has clearly been removed, the 
function approaching a constant at small $x$ as predicted (the erratic 
behaviour at very small $x$ is due to statistical noise). 

For $q=3/2$ all of the numerical simulations have been performed
using the Ising representation, with periodic initial conditions in 
which clusters of $l_0$ `up spins' are placed at uniform intervals 
in a `down spin' background, and occupy a fraction $1/q$ of the sites. 
For $q>2$, the Potts representation has mainly been used, but we expect 
all universal properties, such as exponents and scaling functions, to 
be the same for the two representations. This expectation is confirmed 
by the results for $\theta(q)$ and the scaling function $g(x)$. In order 
to obtain a {\em precise} correspondence between the Ising and the Potts 
representations, however, it is necessary to run the Ising simulations 
with random initial conditions, rather than periodic initial conditions 
which we have mostly employed, and to scale the axes appropriately 
with $q$ to account for the different interval sizes in the two cases. 
This is because the reaction-diffusion representation of the Potts model, 
in which walkers annihilate or coalesce with probabilities which depend 
on $q$, is only an exact representation if the Potts states occur in a 
completely random sequence. This is not true for the periodic initial 
conditions employed in the Ising representation, whose correlations 
spoil the exact correspondence between Ising and Potts simulations. 
Our expectation, then, is that the Ising representation with random 
and periodic initial conditions should give the same {\em scaling 
functions} as the reaction-diffusion implementation of the Potts model, 
but {non-universal amplitudes} will be different for the periodic 
initial conditions.  

\begin{figure}
\narrowtext
\centerline{\epsfxsize\columnwidth\epsfbox{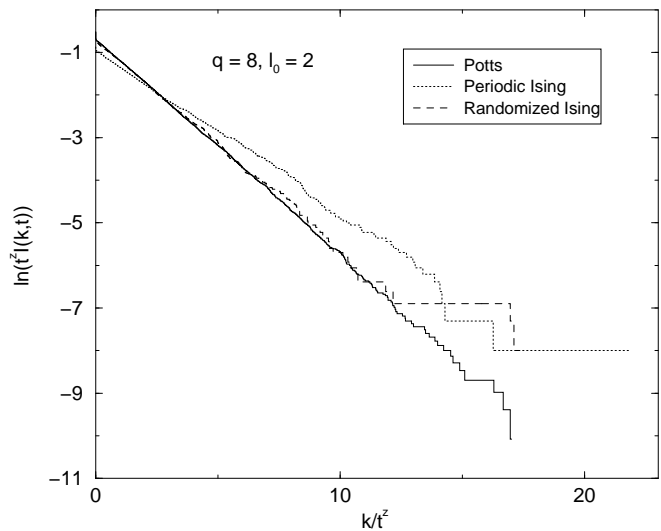}}
\caption{Scaling function for $q=8$, showing equivalence of 
reaction-diffusion (`Potts') and Ising (`Randomized Ising') 
representations, with $z=\theta(8)$, where the axes are 
$qt^zI(k,t)$ and $k/qt^z$ for the Ising data. The Ising data 
for periodic initial conditions (`Periodic Ising') has the same 
functional form but a different amplitude.     
} 
\label{Fig12}
\end{figure}

In order to test the exact equivalence between random Ising and Potts 
simulations we plot, in Figure \ref{Fig12}, the results for the $q=8$ 
model using the Potts (reaction-diffusion) method. The lattice is 
initially all persistent and the initial cluster size is $l_0=2$. 
This corresponds to a sequence of pairs of Potts states in which the 
members of a pair are in same state, but there are no correlations 
between pairs (except that neighboring pairs must be in different 
states). The random initial conditions for the Ising simulation are 
generated from the Potts initial state by setting all pairs in 
Potts state `1' (say) to be up spins and all other pairs to be 
down spins. In the Potts simulations we keep track of all persistent 
sites, while in the Ising simulation we track the persistence of the 
spins initially `up'. The interval sizes (between persistent clusters) 
are then naturally larger by a factor $q$ in the Ising simulations 
relative to the Potts simulations. Figure \ref{Fig12} shows both scaling 
functions in log-linear form, for $t=10^4$, with this factor of $q$ scaled 
out (i.e. $qt^zI(k,t)$ is plotted against $k/qt^z$ for these data). 
The data overlap almost perfectly. The Ising data for a periodic initial 
condition are consistent with the same scaling function (exponential), 
but with a different amplitude.   

\section{Discussion and Summary}
In this paper we have investigated the nature of persistent structures 
in the coarsening dynamics of 1D Potts models. A central concept has 
been the existence of two characteristic length scales, the mean 
separation, $L_w \sim t^{1/2}$, of domain walls (or `walkers') and 
the `persistence length', $L_p \sim t^\theta$, which measures the 
mean separation of persistent sites or clusters. The focus of our 
attention has been the distribution function, $n(k,t)$, for the 
number of intervals (between clusters of persistent sites) of 
length $k$. This distribution has the scaling form (\ref{scaling}), 
with characteristic length scale,  $t^z$, which is the {\em larger} of 
$L_w$ and $L_p$, i.e.\ $z={\rm max}(1/2,\theta)$. Within the 1D Potts 
model both regimes, $\theta < 1/2$ and $\theta > 1/2$, can be 
accessed by varying the number of Potts states, $q$. From the general 
result, (\ref{theta(q)}), we see that $\theta(q)$ is a monotonically 
increasing function with, $q_c = 2.70528\ldots$ marking the boundary 
between the two regimes.  

The regime $\theta > 1/2$ is conceptually simpler. The $q=\infty$ 
limit can been solved exactly (section III), and results for the 
long-time limit of the mean cluster size obtained. The fact that the 
locations of the surviving clusters are statistically independent in 
the scaling regime leads to the result that the scaling function $f(x)$ 
in (\ref{scaling}) is a simple exponential. We have argued that the 
same result should hold for all $\theta > 1/2$, since in this regime 
there are, on average, many walkers between each pair of persistent 
clusters. The {\em relevant} walkers for this argument (those which 
are turning persistent sites in neighboring clusters nonpersistent) 
are uncorrelated since their separations, of order $L_p$, are much 
larger than the typical separation, $L_w$, of neighboring walkers. 
On this basis we expect the asymptotic scaling function $f(x)$ to 
be exponential for all $\theta > 1/2$ (i.e.\ $q>q_c$). The data 
for $q=8$ (Figure \ref{Fig6}) and $q=4$ (Figure \ref{Fig8})  
are consistent with an exponential form for $g(x) = \int_x^\infty f(x)$, 
although the scaling is not perfect and there is a small upturn in the 
scaling function at small scaling variable for $q=4$. We attribute 
these features to an imperfect separation of length scales on the 
time scales achieved in the simulations, and conjecture that the 
true scaling function is exponential for all $\theta>1/2$. 

The case $\theta < 1/2$ is more tricky. In this case the persistent 
clusters outnumber the walkers. The scaling function $f(x)$ is clearly 
{\em not} a simple exponential, though it seems (Figure \ref{Fig10}) to 
have an exponential tail. There is a small-argument singularity of the 
form $f(x) \sim x^{-\tau}$, with $\tau = 2(1-\theta)$. This is related 
to the fractal dimension, $d_f = 1-2\theta$, of the persistent sites: 
$d_f = \tau - 1$. Note that the borderline, $\theta=1/2$, between the 
two regimes occurs at $d_f=0$. The existence of a small-$x$ singularity 
for $\theta < 1/2$ raises the possibility of an alternative scenario 
for $\theta>1/2$, in which the $x^{-\tau}$ singularity, with 
$\tau=2(1-\theta)$, persists for $\theta > 1/2$, where it becomes an 
integrable singularity. The small-$x$ singularity in $g(x)$ would 
then take the form of a cusp: $g(x) = g(0) - Ax^{2\theta - 1} + \cdots$. 
We have not been able to rule out this scenario from the data, but think 
it unlikely for the reasons given elsewhere in this paper. Further 
insight could be obtained if it were possible to perform an expansion 
around the soluble $q=\infty$ limit to first order in $1/q$, but we 
leave this as a challenge for the future.  

We conclude by discussing briefly the possibility of the existence of 
these two qualitatively different regimes in spatial dimension $d \ge 2$. 
First we generalize the result (\ref{df}), relating $d_f$ and $\theta$,  
to any dimension.  Starting from (\ref{both}), the result $F(x) \sim 
x^{-2\theta}$ follows generally, and the number of persistent sites 
within a distance $R$ of a given site is estimated as  
$\int_0^R r^{d-1}dr\,r^{-2\theta} \sim R^{d_f}$, where $d_f = d-2\theta$. 
Generalizing still further, if the coarsening exponent is $\phi$, rather 
than $1/2$, the result is 
\begin{equation}
d_f = d - \theta/\phi 
\end{equation}
(the $d=2$ version of this result is given in \cite{Jain}, based on the 
same reasoning). Clearly this result requires $\theta \le d\phi$, since 
$d_f$ cannot be negative. If this inequality is violated, as in the 1D 
Potts model with $\theta>1/2$, the persistent sites no longer have a 
fractal structure but become point-like objects, with mean density 
$\sim t^{-\theta} \ll t^{-d\phi} \sim L_c^{-d}$, where $L_c \sim t^\phi$ 
is the coarsening length scale. Cases where $\theta < d\phi$ are easy 
to find, e.g.\ in the coarsening of the 2D Ising model \cite{p2}, the 
2D diffusion equation \cite{p3}, or the time-dependent Ginzburg-Landau 
equation in 2D. These all exhibit fractal persistent structures 
with the expected fractal dimension \cite{Jain,Cornell}. It would be 
interesting to look for examples, in addition to the 1D Potts model, 
where one can have $\theta > d\phi$.  

\section{Acknowledgments}
This work was supported by EPSRC (UK).

\section{Appendix}
In this Appendix we compute the probability that a cluster of initial 
size $l_0$ survives to time $t$, and the mean size of surviving clusters, 
for the process $A + A \to A$. 

First note that initially there is a random walker at each end of the 
domain. At each time step the walkers move independently left or right 
with probability $1/2$, so we can treat each walker independently. 
Consider, therefore, a single walker moving at discrete time steps on a 
discrete 1D lattice, starting, at time $t=0$, at position $r$. 
Let the `origin' be the point $r=0$, and let $P_r(t)$ be the probability 
that the walker has not yet reached the origin at time $t$. Clearly, 
\begin{eqnarray}
P_1(t) & = & \frac{1}{2}\,P_2(t-1) \nonumber \\
P_r(t) & = & \frac{1}{2}\,[P_{r-1}(t-1) + P_{r+1}(t-1)],\ \ \ r \ge 2\ .
\label{recurrence}
\end{eqnarray}
We are interested only in the limit $t \to \infty$. In this limit, we 
know from standard random walk theory that every $P_r(t)$ decays like 
$t^{-1/2}$, with an $r$-dependent amplitude. To leading order in $t^{-1/2}$, 
therefore, the $t$-dependence drops out of equations (\ref{recurrence}), 
which then become equations for the amplitudes. By inspection, the solution 
in this regime is 
\begin{equation}
P_r(t) = rP_1(t)\ .
\label{leading}
\end{equation} 

Now consider a walker starting immediately to the right of a cluster of 
$l_0$ persistent sites. The probability that after $t$ steps the walker 
has jumped over exactly $r$ of these (making them nonpersistent) is 
$P_{r+1}(t) - P_r(t) = P_1(t)$, where the final result follows from 
(\ref{leading}), to leading order for large $t$. The same result holds 
for a walker starting immediately to the left of the cluster.  
The probability that $l$ sites remain persistent after time $t$ is 
$P_1(t)^2$ times the number of ways of partitioning the cluster of length 
$l_0$ into 3 sections, with the central section of length $l$ (and zero 
lengths are allowed for the outer sections). This number is $l_0-l+1$. 
So the probability of the cluster surviving and having length $l$ is 
\begin{equation}
p_{l_0}(l,t) = (l_0 - l + 1)\,P_1(t)^2\ ,
\end{equation}
a generalization of (\ref{pl0}) to the discrete system. 

The survival probability of the cluster is
\begin{equation} 
p_{surv}(t) = \sum_{l=1}^{l_0} p_{l_0}(l,t) 
= \frac{1}{2}l_0(l_0+1)\,P_1(t)^2\ ,
\end{equation}
while the mean cluster size is 
\begin{equation}
\langle l \rangle_\infty = \frac{\sum_{l=1}^{l_0}l\, p_{l_0}(l,t)}
{\sum_{l=1}^{l_0} p_{l_0}(l,t)} = \frac{(l_0+2)}{3}\ .
\label{l_av}
\end{equation} 

\end{multicols}

\end{document}